# How simple rules determine pedestrian behavior and crowd disasters

Mehdi Moussaïd[1, 2, 3,*], Dirk Helbing[2,4], and Guy Theraulaz[1,3]

[1] Centre de Recherches sur la Cognition Animale, UMR-CNRS 5169, Université Paul Sabatier, Bât 4R3, 118 Route de Narbonne, 31062 Toulouse cedex 9, France. Phone: +33 5 61 55 64 41 ; Fax: +33 5 61 55 61 54.

[2] ETH Zurich, Swiss Federal Institute of Technology, Chair of Sociology, CLU E1, Clausiusstrasse 50, 8092 Zurich, Switzerland. Phone +41 44 632 88 81; Fax +41 44 632 17 67.

[3] CNRS, Centre de Recherches sur la Cognition Animale, F-31062 Toulouse, France

[4] University of Oxford, Nuffield College, New Road, Oxford OX1 1NF, Great Britain

* Corresponding author: moussaid@cict.fr

## *Classification*

Social Sciences/Psychological and Cognitive Sciences

Physical Sciences/Applied Physical Sciences

*Abstract*

With the increasing size and frequency of mass events, the study of crowd disasters and the simulation of pedestrian flows have become important research areas. Yet, even successful modeling approaches such as those inspired by Newtonian force models are still not fully consistent with empirical observations and are sometimes hard to calibrate. Here, a novel cognitive science approach is proposed, which is based on behavioral heuristics. We suggest that, guided by visual information, namely the distance of obstructions in candidate lines of sight, pedestrians apply two simple cognitive procedures to adapt their walking speeds and directions. While simpler than previous approaches, this model predicts individual trajectories and collective patterns of motion in good quantitative agreement with a large variety of empirical and experimental data. This includes the emergence of self-organization phenomena, such as the spontaneous formation of unidirectional lanes or stop-and-go waves. Moreover, the combination of pedestrian heuristics with body collisions generates crowd turbulence at extreme densities—a phenomenon that has been observed during recent crowd disasters. By proposing an integrated treatment of simultaneous interactions between multiple individuals, our approach overcomes limitations of current physics-inspired pair interaction models. Understanding crowd dynamics through cognitive heuristics is therefore not only crucial for a better preparation of safe mass events. It also clears the way for a more realistic modeling of collective social behaviors, in particular of human crowds and biological swarms. Furthermore, our behavioral heuristics may serve to improve the navigation of autonomous robots.

## *Introduction*

Human crowds display a rich variety of self-organized behaviors that support an efficient motion under everyday conditions (1-3). One of the best-known examples is the spontaneous formation of uni-directional lanes in bi-directional pedestrian flows. At high densities, however, smooth pedestrian flows can break down, giving rise to other collective patterns of motion such as stop-and-go waves and crowd turbulence (4). The latter may cause serious trampling accidents during mass events. Finding a realistic description of collective human motion with its large degree of complexity is therefore an important issue.

Many models of pedestrian behavior have been proposed in order to uncover laws underlying crowd dynamics (5-8). Among these, physics-based approaches are currently very common. Well-known examples are fluid-dynamic (9) and social force models (1, 7, 8, 10), which are inspired by Newtonian mechanics. The latter describe the motion of pedestrians by a sum of attractive, repulsive, driving, and fluctuating forces reflecting various external influences and internal motivations. However, even though physics-inspired models are able to reproduce some of the observations quite well, there are still a number of problems. Firstly, it is becoming increasingly difficult to capture the complete range of crowd behaviors in one single model. Recent observations have required extensions of previous interaction functions, which have led to quite sophisticated mathematical expressions that are relatively hard to calibrate (10). Secondly, these models are based on the superposition of binary interactions. For example, in a situation where an individual A is facing three other individuals B, C and D, the behavior of A is given by an integration of the interaction effects that the three individuals would separately have on A in the absence of the others. However, this raises many theoretical issues, such as how to integrate the binary interactions (e.g. to sum them up, average over them, or combine them non-linearly), how to determine influential neighbors (e.g. the closest $N$ individuals or those in a certain radius $R$), and how to weight their influence (e.g. when located to the side or behind the focal pedestrian) (6, 11, 12).

Here, we propose instead a novel cognitive science approach based on behavioral heuristics, which overcomes the above problems. Heuristics are fast and simple cognitive procedures that are often used when decisions have to be made under time pressure, or overwhelming information (13, 14). Let us illustrate this by the example of a player trying to catch a ball, which may be modeled in at least two ways: either, an attraction force can be used to describe the player's motion toward the estimated landing point of the ball, or the process can be described by a so-called "gaze heuristic". This consists of visually fixating on the ball and adjusting the position such that the gazing angle remains constant. Both methods predict similar behavior, but the heuristic approach is simpler and more plausible.

Heuristics have also successfully explained decision-making in a variety of situations such as the investment behavior at stock markets or medical diagnosis in emergency situations (13). Modeling the *collective* dynamics of a social system with many interacting individuals would be a promising new approach. However, is it possible to apply a heuristics approach to pedestrian motion as well, given the wealth of different crowd dynamics patterns that have been observed?

In this work, we show that two simple heuristics based on visual information can in fact describe the motion of pedestrians well, and that most properties observed at the crowd level follow naturally from them. Moreover, the combination of pedestrian heuristics with body collisions reproduces observed features of crowd disasters at extreme densities.

## *Model*

The elaboration of a cognitive model of pedestrian behavior requires two crucial questions to be addressed: (a) "*What kind of information is used by the pedestrian?*" and (b) "*How is this information processed to adapt the walking behavior?*". With regard to the first question, past studies have shown that vision is the main source of information used by pedestrians to control their motion (15-17). Accordingly, we will start with the representation of the visual information of pedestrians. To answer the second question, we propose two heuristics based on this visual information, which determine the desired walking directions $\alpha_{des}$ and desired walking speeds $v_{des}$ of pedestrians. Finally, we assume that pedestrians are continuously adapting their current walking behavior to match their desired behavior with a relaxation time $\tau$ of 0.5 seconds (see Fig. S1 in the SI). This has been confirmed under controlled laboratory conditions (10).

*Representation of visual information*

In our model, each pedestrian *i* is characterized by its current position $\vec{x}_i$ and speed $\vec{v}_i$. For simplicity, we represent the projection of a pedestrian's body on the horizontal plane by a circle of radius $r_i = m_i/160$, where $m_i$ is the mass of pedestrian *i* (e.g. uniformly distributed in the interval [60kg 100kg]). Each pedestrian is additionally characterized by his or her comfortable walking speed $v_i^0$, and his or her destination point $O_i$, namely the place in the environment he or she wants to reach, such as the exit door of a room, or the end of a corridor. Finally, the vision field of pedestrian *i* ranges to the left and to the right by $\phi$ degrees with respect to the line of sight $\vec{H}_i$.

Past studies have shown that walking subjects can estimate the time to collision with surrounding obstacles thanks to specialized neural mechanisms at the retina and brain levels (18, 19). Accordingly, we represent the pedestrian's visual information as follow: For all possible directions $\alpha$ in $[-\phi, \phi]$ (with a reasonable angular resolution), we compute the distance to the first collision $f(\alpha)$, if pedestrian *i* moved in direction $\alpha$ at speed $v_i^0$, taking into account the other pedestrians' walking speeds and body sizes. If no collision is expected to occur in direction $\alpha$, $f(\alpha)$ is set to a default maximum value $d_{max}$, which represents the "horizon distance" of pedestrian *i* (see **Fig. 1**).

*Formulation of the cognitive heuristics*

The first movement heuristic concerns the relative angle $\alpha_{des}$ of the chosen walking direction compared to the line of sight. Empirical evidence suggests that pedestrians seek an unobstructed walking direction, but dislike deviating too much from the direct path to their destination (16, 17). A trade-off therefore has to be found between avoiding obstacles and minimizing detours from the most direct route. Accordingly, our first heuristic is: "*A pedestrian chooses the direction $\alpha_{des}$ that allows the most direct path to destination point $O_i$, taking into account the presence of obstacles*". The chosen direction $\alpha_{des}(t)$ is computed through the minimization of the distance $d(\alpha)$ to the destination:

$$d(\alpha) = d_{max}^2 + f(\alpha)^2 - 2d_{max} f(\alpha) \cos(\alpha_0 - \alpha).$$

Here, $\alpha_0$ is the direction of the destination point.

The second heuristic determines the desired walking speed $v_{des}(t)$. Since a time period $\tau$ is required for the pedestrian to stop in the case of an unexpected obstacle, pedestrians should compensate for this delay by keeping a safe distance (20). Therefore, we formulate the second heuristic as follows: *"A pedestrian maintains a distance from the first obstacle in the chosen walking direction that ensures a time to collision of at least $\tau$."* In other words, the speed $v_{des}(t)$ is given by: $v_{des}(t) = \min(v_i^0, d_h/\tau)$, where $d_h$ is the distance between pedestrian $i$ and the first obstacle in the desired direction $\alpha_{des}$ at time $t$. The vector $\vec{v}_{des}$ of the desired velocity points in direction $\alpha_{des}$ and has the norm $\|\vec{v}_{des}\| = v_{des}$. The change in the actual velocity $\vec{v}_i$ at time $t$ under normal walking conditions is given by the acceleration equation $d\vec{v}_i/dt = (\vec{v}_{des} - \vec{v}_i)/\tau$.

*Effect of body collisions*
In cases of overcrowding, physical interactions between bodies may occur, causing unintentional movements that are not determined by the above heuristics. Indeed, at extreme densities, it is necessary to distinguish between the *intentional* avoidance behavior of pedestrians adapting their motion according to perceived visual cues, and *unintentional* movements resulting from interaction forces caused by collision with other bodies. We have therefore extended the above description by considering physical contact forces

$$\vec{f}_{ij} = kg(r_i + r_j - d_{ij})\vec{n}_{ij},$$

where $g(x)$ is zero if the pedestrians $i$ and $j$ do not touch each other, and otherwise equals the argument $x$. $\vec{n}_{ij}$ is the normalized vector pointing from pedestrian $j$ to $i$, and $d_{ij}$ is the distance between the pedestrians' centers of mass (1). The physical interaction with a wall $W$ is represented analogously by a contact force $\vec{f}_{iW} = kg(r_i - d_{iW})\vec{n}_{iW}$, where $d_{iW}$ is the distance to the wall $W$ and $\vec{n}_{iW}$ is the direction perpendicular to it.

The resulting acceleration equation reads $d\vec{v}_i/dt = (\vec{v}_{des} - \vec{v}_i)/\tau + \sum_j \vec{f}_{ij}/m_i + \sum_W \vec{f}_{iW}/m_i$ and is solved together with the usual equation of motion $d\vec{x}_i/dt = \vec{v}_i$, where $\vec{x}_i(t)$ denotes the location of pedestrian $i$ at time $t$. In contrast to social force models, however, the interaction terms $\vec{f}_{ij}$ and $\vec{f}_{iW}$ are non-zero *only* in extremely crowded situations, but not under normal walking conditions.

## *Results*

The combination of behavioral heuristics with contact forces accounts for a large set of complex collective dynamics. In the following section, we will first validate the model at the individual level, and then explore the model predictions in a collective context for uni- and bi-directional flows.

*Individual trajectories*

Firstly, we tested the model in the context of simple interaction situations involving two pedestrians avoiding each other. In a series of laboratory experiments, we tracked the motion of pedestrians in two well-controlled conditions: (a) passing a pedestrian standing in the middle of a corridor, and (b) passing another pedestrian moving in the opposite direction (see Material and Methods) (10). The model predicts individual avoidance trajectories that agree very well with the experimentally observed trajectories under both conditions (**Fig. 2**).

*Collective patterns of motion*

Next, we explored the model predictions in a collective context. For bidirectional traffic in a street, assuming random initial positions of pedestrians, we find that flow directions separate spontaneously after a short time, as empirically observed (see Fig. S2 in the SI). This reflects the well-known lane formation phenomenon (2), which is a characteristic property of crowd dynamics.

We have also investigated the influence of pedestrian density on unidirectional flows. The velocity-density relation predicted by the model agrees well with empirical data (21) **(Fig. 3a)**. Furthermore, when the density exceeds critical values, our model shows transitions from smooth flows to stop-and-go waves and "crowd turbulence", as has been observed before crowd disasters (4). **Figure 3c** depicts typical space-time diagrams for simulations at various density levels, displaying a smooth, laminar flow at low density (regime 1), but stop-and-go waves at higher densities (regimes 2 and 3). These result from the amplification of small local perturbations in the flow due to coordination problems when competing for scarce gaps (22): When the density of pedestrians is high enough, such perturbations trigger a chain reaction of braking maneuvers, resulting in backward moving waves. This is illustrated by the significant correlation between the local speed at positions $x_1$ and $x_2 = x_1 - X$ after a certain time lag $T$ **(Fig. 3b)**. In particular, the model allows us to estimate the backward propagation speed of the wave (approximately 0.6 m/s) and the density interval where stop-and-go waves occur (at occupancy levels between 0.4 and 0.65, i.e. 40 to 65% spatial coverage).

At even higher densities, physical interactions start to dominate over the heuristic-based walking behavior (see **inset of Fig. 3a).** As the interaction forces in the crowd add up, intentional movements of pedestrians are replaced by unintentional ones. Hence, the well-coordinated motion among pedestrians suddenly breaks down, particularly around bottlenecks (see Figs. 4a and S4 in the SI). This results in largely fluctuating and uncontrollable patterns of motion, called "crowd turbulence". A further analysis of the phenomenon reveals areas of serious body compression occurring close to the bottleneck (see **Fig. 4a**). The related, unbalanced pressure distribution results in sudden stress releases and earthquake-like mass displacements of many pedestrians in all possible directions (4) (see **Figs. 4b and 4c**). The distribution of displacements predicted by the model is well approximated by a power law with exponent $1.95 \pm 0.09$. This is in excellent agreement with detailed evaluations of crowd turbulence during a crowd disaster that happened to be recorded by a surveillance camera (4).

## *Discussion*

The greater explanatory power of our heuristics-based modeling, as demonstrated through comparison with different empirical and experimental data (see the overview in Table S1), suggests a *paradigm shift* from physics-inspired binary interaction models to an integrated treatment of multiple interactions, which are typical for social interactions in human crowds or animal swarms (23-28). Without requiring additional assumptions, our approach overcomes various issues related to the combination of multiple binary interactions (6, 11). Our model treats a pedestrian's reaction to his or her visually perceived environment in an integrated way rather than reducing it to a superposition of pair interactions. Instead of being *repelled* by their neighbors, as was assumed in previous particle models, individuals actively *seek a free path* through the crowd. The combined effect of neighboring individuals is implicitly included in the representation of a pedestrian's visual field. Our new model therefore correctly handles situations in which pedestrians are hidden or outside the field of view. Finally, high-density and life-threatening situations can be studied by combining heuristics-based movement resulting from visual perception of the environment with unintentional displacements due to physical forces resulting from unavoidable collisions with other bodies. In doing so, the emergence of crowd turbulence in panic situations can be reproduced as well.

Understanding pedestrian heuristics and the emergence of complex crowd behavior is a crucial step towards a more reliable description and prediction of pedestrian flows in real-life situations. Our heuristics-based model therefore has important practical applications, such as the improvement of architectures and exit routes, as well as the organization of mass events. In addition, the vision-based treatment of the pedestrian heuristics appears to be particularly suited to the study of evacuation conditions with reduced visibility (e.g. escaping from a smoke-filled room) (2, 29).

In future, further evidence for our cognitive, heuristics-based model could be collected by using eye-tracking systems (30) to determine the visual cues followed by pedestrians. Our approach also opens new perspectives in other research areas. In the field of autonomous robotics, for example, the model may serve to improve navigation in complex dynamic environments. This is particularly relevant for swarms of mobile robots (31). In fact, navigation and collision-avoidance concepts of multi-robot systems have often been inspired by human behavior (32, 33). The simplicity of our new approach and its visual information input will support resource-efficient designs. We also expect that our heuristics-based approach will inspire new models of collective human behavior such as group decision making (34) and certain social activity patterns (35, 36), where the occurrence of simultaneous interactions between multiple individuals matters.

## *Material and Methods*

**Experimental setup.** The controlled experiments shown in Fig. 2 were conducted in 2006 in Bordeaux (France). The experimental corridor of 7.88m length and 1.75m width was equipped with a three-dimensional tracking system, which consisted of three digital cameras (SONY DCR-TRV950E) mounted at the corners of the corridor. The reconstruction of the positions was made on the basis of the digital movies encoded at 12 frames per seconds with the help of software developed in our team. The trajectories were smoothed over a time window of 10 frames. A total of 40 participants agreed to participate in the experiment and were naïve to its purpose. Pairs of pedestrians were randomly matched and performed approximately twenty replications of the two following conditions: (1) One subject was instructed to stand still in the middle of the corridor, while the other one was instructed to walk from one end of the corridor to the other and had to evade the standing pedestrian. (2) Starting from opposite ends of the corridor, both subjects were instructed to walk toward the other end after the starting signal. A total of 148 and 123 trajectories were reconstructed for conditions 1 and 2 respectively.

**Definition of local variables.** The simulation results presented in the main text were analyzed by measuring the local speed, local "pressure" and local compression coefficients (4). The local speed $V(x,t)$ at place $x$ and time $t$ (used in Fig. 3C) was defined as

$$V(x,t) = \frac{\sum_i \|\vec{v}_i\| f(d_{ix})}{\sum_i f(d_{ix})},$$

where $d_{ix}$ is the distance between $x$ and pedestrian $i$. In this definition, $f(d)$ is a Gaussian distance-dependent weight function defined as:

$$f(d) = \frac{1}{\pi R^2} \exp(-d^2/R^2),$$

where $R$ is a measurement parameter. The value $R=0.7m$ provides a reasonably precise evaluation of the local speed. The local body compression coefficient $C(x)$ (used in Fig. 4A) was computed in a way analogous to the local speed, setting

$$C(x,t) = \frac{\sum_i C_i(t) f(d_{ix})}{\sum_i f(d_{ix})},$$

and $C(x) = \langle C(x,t) \rangle_t$, where the brackets denote an average over time. The body compression $C_i(t)$ of a pedestrian $i$ is the sum of the contact forces $\vec{f}_{ij}$ applied to pedestrian $i$:

$$C_i(t) = \sum_j \|\vec{f}_{ij}(t)\|$$

Finally, the critical zones identified in Fig. 4B are given by the "crowd pressure" $P(x) = \rho(x) Var(V(x,t))$ defined in Ref. (4), i.e. the pressure corresponds to the average local density $\rho(x) = \sum_i f(d_{ix})$ times the local speed variance at place $x$.

## *Acknowledgments*

We are grateful to A. Johansson, S. Garnier, M. Moreau, D. Boyer, J. Gautrais, and H. Chaté for inspiring discussions and to Suzy Moat for language editing. We thank A. Campo, F. Ducatelle and the IDSIA research group in Manno-Lugano, Switzerland for useful suggestions. M.M. was supported by a joint doctoral-engineer fellowship from ETH Zurich and CNRS. This study was supported by grants from the CNRS (Concerted Action: Complex Systems in Human and Social Sciences), the University Paul Sabatier (Aides Ponctuelles de Coopération) and the PEDIGREE project (Grant No. ANR-08-SYSC-015).

*Figures*

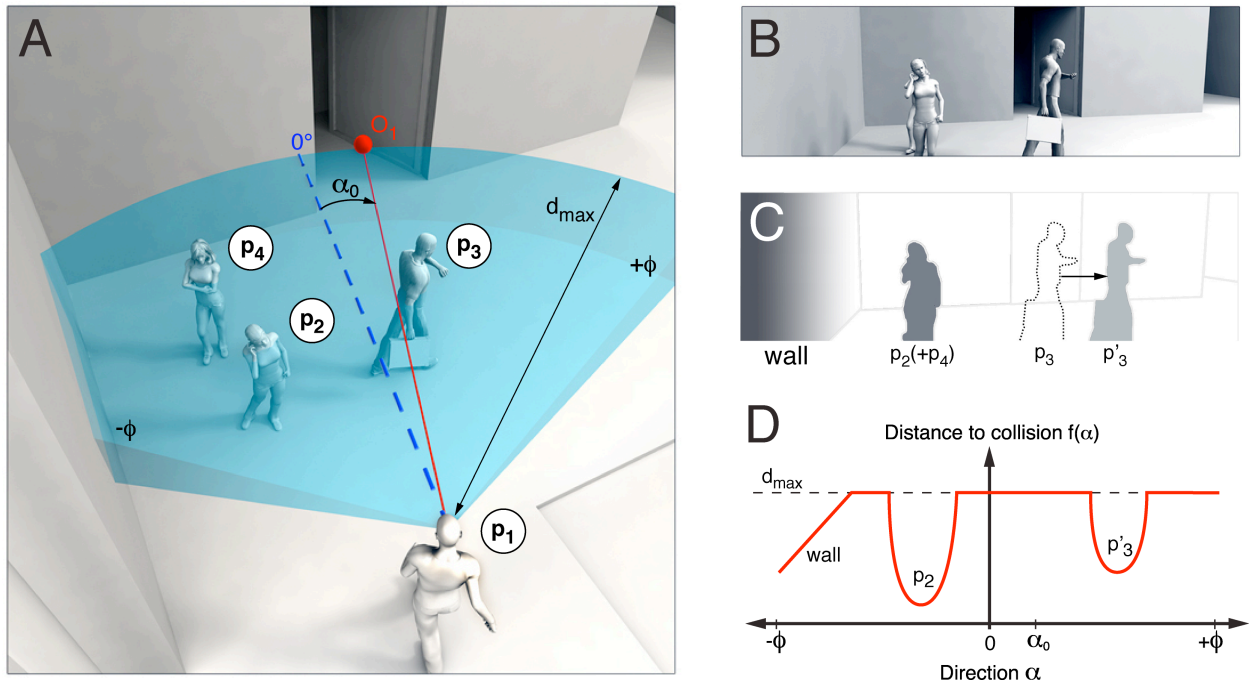

**Figure 1:** **(A)** Illustration of a pedestrian $p_1$ facing three other subjects and trying to reach the destination point $O_1$ marked in red. The blue dashed line corresponds to the line of sight. **(B)** Illustration of the same situation, as seen by pedestrian $p_1$. **(C)** Abstraction of the scene by a black and white visual field. Here, darker areas represent a shorter collision distance. **(D)** Graphical representation of the function $f(\alpha)$ reflecting the distance to collision in direction $\alpha$. The left-hand side of the vision field is limited by a wall. Pedestrian $p_4$ is hidden by pedestrian $p_2$ and, therefore, not visible. Pedestrian $p_3$ is moving away, so a collision would occur in position $p'_3$, but only if $p_1$ moved towards the right-hand side.

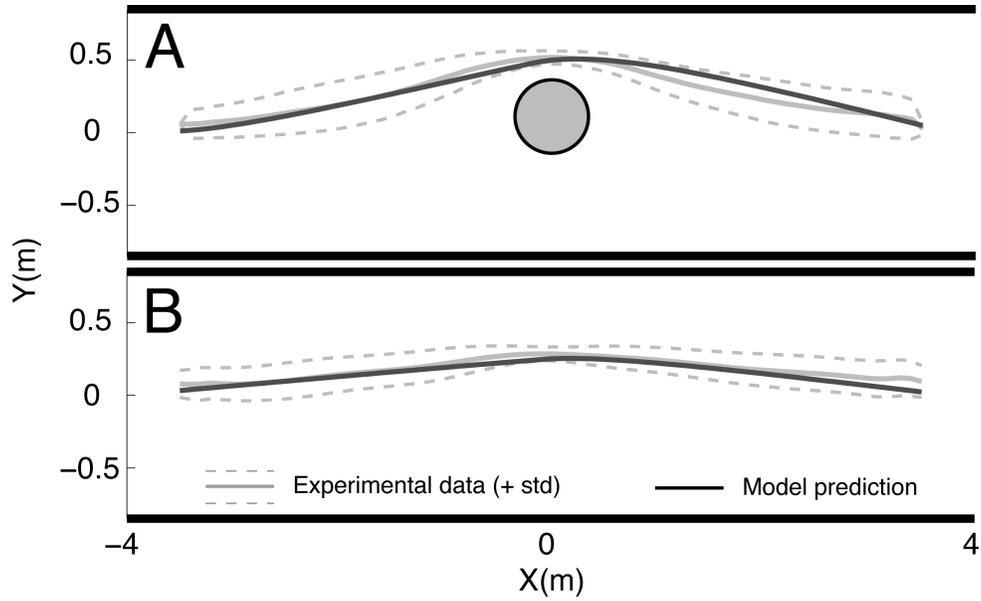

**Figure 2:** Results of computer simulations for the heuristic pedestrian model (solid black lines) as compared to experimental results (light grey) during simple avoidance maneuvers in a corridor of 7.88m length and 1.75m width (data from Ref. (10)). (**A**) Average trajectory of a pedestrian passing a static individual standing in the middle of the corridor (*N*=148 replications). (**B**) Average trajectory of a pedestrian (solid light grey line) passing another individual moving into the opposite direction (*N*=123 replications). Dashed lines indicate the standard deviation of the average trajectory. Pedestrians are moving from left to right. The computer simulations were conducted in a way that reflected the experimental conditions. The model parameters are $\tau=0.5$s, $\phi=75°$, $d_{max}=10$m, $k=5.10^3$, $v_i^0=1.3$ m/s.

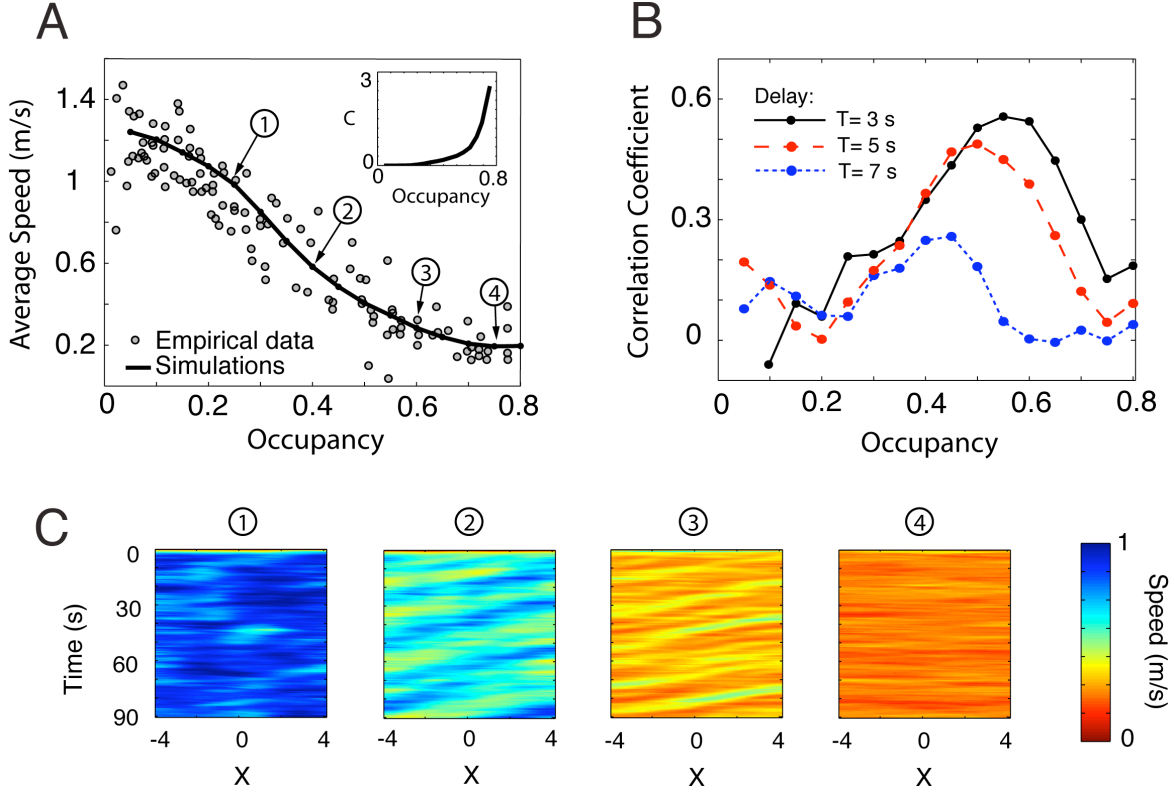

**Figure 3:** Evaluation of different kinds of collective dynamics resulting for unidirectional flows in a street of length $l$=8m and width $w$=3m. The total number of pedestrians is varied from 6 to 96, assuming periodic boundary conditions. **(A)** Velocity-density relation, determined by averaging over the speeds of all pedestrians for 90 seconds of simulation. The occupancy corresponds to the fraction of area covered by pedestrian bodies. Our simulation results (black curve) are well consistent with empirical data (dots), which were collected in real-life environments (21). The inset indicates the average body compression $C = \langle C_i(t) \rangle_{i,t}$ (see Materials and Methods), where the brackets indicate an average over all pedestrians $i$ and over time $t$. **(B)** Correlation coefficient between the average local speeds $V(x,t)$ and $V(x-X, t+T)$, measuring the occurrence of stop-and-go waves (see Material and Methods for the analytical definition of the local speed). Here, the value of $X$ is set to 2m. The increase at intermediate densities indicates that speed variations at positions $x$ and $x-X$ are correlated for an assumed time delay $T$ of 3 seconds. Significant p-values for the correlation coefficient are found for occupancies between 0.4 and 0.65, indicating the boundaries of the stop-and-go regime (Fig. S3). **(C)** Typical space-time diagrams at four density levels, representing different kinds of collective motion. The color coding indicates the local speed values along the street (where pedestrians move from left to right). At occupancy level 1, the diagram displays a smooth, laminar flow with occasional variations in speed. For occupancy levels 2 and 3, stop-and-go waves appear, as they have been empirically observed at high densities [see Fig. 2a in Ref. (4)]. At occupancy level 4, the average traffic flow is almost zero, but turbulent fluctuations in the flow occur (see Fig. 4). The underlying model parameters are $\tau$=0.5s, $\phi$=45°, $d_{max}$=8m, $k=5.10^3$. The desired speed $v_i^0$ was chosen according to a normal distribution with mean value 1.3 m/s and standard deviation 0.2.

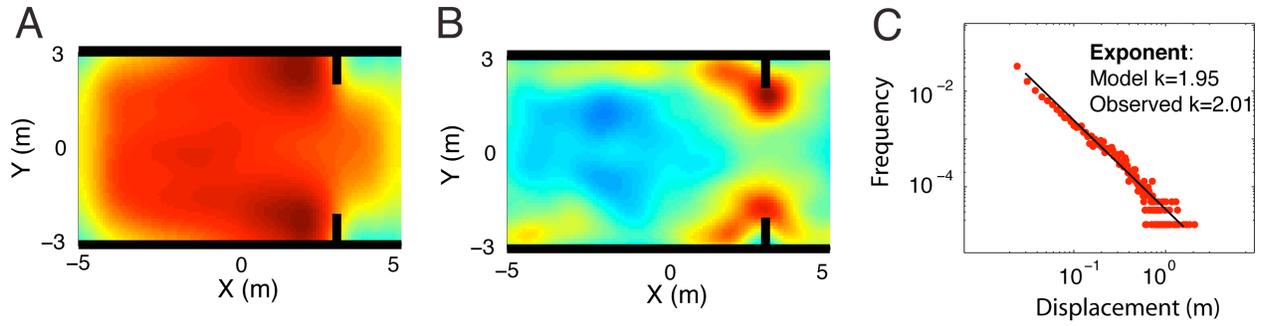

**Figure 4:** Characterization of turbulent flows in front of a bottleneck for an occupancy value of 0.98. (For the analysis of a turning corridor as in the Love Parade disaster in Duisburg in 2010 see Fig. S4 in the SI). **(A)** The local body compression $C(\vec{x})$ reveals two critical areas of strong compression in front of the bottleneck (shown in red). **(B)** Analyzing the "crowd pressure" (defined as local density times the local velocity variance, see Materials and Methods) reveals areas with a high risk of falling (in red), indicating the likelihood of a crowd disaster (4). **(C)** Distribution of displacements (i.e. location changes between two subsequent stops, defined by speeds with $\|\vec{v}_i\| < 0.05 m/s$). The double logarithmic representation reveals a power law with slope $k = -1.95 \pm 0.09$, in good agreement with empirical findings [see Fig. 3e in Ref. (4), where the slope is $k = -2.01 \pm 0.15$]. The local speed, local pressure and local compression coefficients are defined in the Material and Methods section. The above results are based on simulations of 360 pedestrians during 240 seconds in a corridor of length $l$=10m and width $w$=6m, with a bottleneck of width 4 m, assuming periodic boundary conditions.